\documentclass[12pt,preprint]{article}
\usepackage{graphics}
\usepackage{epsfig}
\usepackage{bm}
\usepackage{amsmath}
\usepackage{amssymb}

\title {Neutrino flavour relaxation \\or\\ neutrino oscillations?}
\author{ I.N. Machulin, S.V. Tolokonnikov
\thanks{e-mail: machulin@in2p3.fr, tolkn@mbslab.kiae.ru}}
\date{\today}

\begin{document}
\newcommand{\beq}{\begin{equation}}
\newcommand{\eeq}{\end{equation}}

\begin{titlepage}

\maketitle
\thispagestyle{empty}
\begin{center}
{\it Russian Research Centre "Kurchatov Institute",\\
123182, Kurchatov sq. 1, Moscow, Russia }
\end{center}
\vspace{12pt}
\begin{abstract}
    We propose the new mechanism of neutrino flavour relaxation
to explain the experimentally observed changes of initial neutrino
flavour fluxes. The test of neutrino relaxation hypothesis is
presented, using the data of modern reactor, solar and accelerator
experiments. The final choice between the standard neutrino
oscillations and the proposed neutrino flavour relaxation
model can be done in future experiments.
\end{abstract}

{\quad \small PACS: 14.60.-z;14.60.Pq;76.20.+q }

\end{titlepage}

    Now the phenomena of changes in initial neutrino flavour flux
is observed in different neutrino experiments. The SNO experiment
\cite{1} evidently detects only 1/3 part of initial electron
neutrino flux from the Sun and 2/3 part of muon and(or) tau
neutrino fluxes. The Kamland reactor experiment \cite{2} detected
only 61\% of expected electron antineutrino events from different
reactors at mean distance of 180 km. Convincing evidence of
initial neutrino flavour flux changes is observed also in Super-K
\cite {3} and MACRO \cite{4} atmospheric neutrino data and K2K
\cite{5} accelerator experiment with muon neutrinos.

    The standard common way of interpreting these results lies in
neutrino oscillation hypothesis, first proposed by Bruno
Pontecorvo \cite {6} and developed in further works \cite {7}.

    Here we discuss the alternative mechanism of neutrino flavour
relaxation, which can also describe the observed changes of
initial neutrino flavour fluxes with distance.

    The proposed model is similar to the mechanism \cite {8} of spin
relaxation in random fluctuating magnetic field $\bf B$ with zero
average $<{\bf B}(t)> = 0$ and mean square fluctuating field value
$<{\bf B}^2(t)> \ne 0$. The spin relaxation process is described
by W.Pauli master equation \cite {9}.

    Let us assume the existence of some small random fluctuating vacuum
field $\widehat V$, causing the transitions between different
lepton flavours. Such field $\widehat V$ can have mean zero value,
with $<{\widehat V}^2(t)>$ different from 0 and $(<{\widehat
V}^2>)^{1/2}
> m_{\nu_e},\, m_{\nu_\mu},\, m_{\nu_\tau}$. Interactions of neutrino
 flavours with the vacuum field should lead to flavour relaxation process.

    The time evolution of the
neutrino states are governed by the Shrodinger equation:

\beq
 i\frac d{dt} |\,\nu(t)> = {\widehat H}(t)|\,\nu(t)>\,, \eeq

where $|\,\nu(t)>$ is the neutrino vector of state and
 H(t) is the time-dependent Hamiltonian of the system, which form
 depends on in what basis it is given.

   In flavour basis the total Hamiltonian in the random fluctuating
vacuum field for rest reference frame can be written as

\beq
    {\widehat H}_f(t)={\widehat H}_m + {\widehat V}_f(t)\,,
\eeq

where

\beq
   {\widehat H}_m=\left(\begin{array}{ccc} m_1 & 0 & 0 \\
 0 & m_2 & 0 \\0 & 0 & m_3 \end{array}\right)\quad  {\rm  and} \quad  {\widehat V}_f(t)=
 \left(\begin{array}{ccc} 0 & V_{e \mu}(t) & V_{e \tau}(t) \\
 V_{\mu e}(t) & 0 & V_{\mu \tau}(t) \\ V_{\tau e}(t) & V_{\tau \mu}(t) & 0 \end{array}
\right)\,.
\eeq

   Here $H_m$ is the free Hamiltonian in mass basis.
    Note, that in this model the neutrino flavour states are assumed to
 be the same as their massive states (or in other words the mixing
  mass matrix is diagonal).

$V_{\alpha \alpha'}$ $ ({\alpha= e, \mu,\tau})$ is the vacuum field potential
with mean value $<V_{\alpha \alpha'}(t)> = 0$ and $<{(V_{\alpha \alpha'}(t))}^2> \ne 0$.

    The neutrino flavour evolution in time can be obtained from Eq. (1-3), using the
density matrix approach \cite {10} and is given by Pauli master equation:

\beq
 \frac d{dt}
  \left(\begin{array}{c} n_{\nu_e} \\ n_{\nu_\mu} \\ n_{\nu_\tau}\end{array}\right)
 =\left(\begin{array}{ccc} -(W_{\mu e}+W_{\tau e}) & W_{e \mu} & W_{e \tau} \\
 W_{\mu e} & -(W_{e \mu}+ W_{\tau \mu})& W_{\mu \tau} \\
 W_{\tau e} & W_{\tau \mu} & -(W_{e \tau}+ W_{\mu \tau}) \end{array}\right)
 \left(\begin{array}{c} n_{\nu_e} \\ n_{\nu_\mu} \\
 n_{\nu_\tau}\end{array}\right)\,,
\eeq

where $n_\alpha(t)\, (\alpha=\nu_e, \nu_\mu, \nu_\tau)$ are the probability of
observing neutrino with electron, muon or tau flavour, $W_{\alpha\alpha`}$
corresponds to the neutrino transition rates  from flavour $\alpha`$ to
flavour $\alpha$ and
\beq \sum_{\alpha=\nu_e, \nu_\mu, \nu_\tau} n_\alpha(t)=1\,.\nonumber\ \eeq

    The general solution of Eq. (4) is given by the sum of
two exponents and constant:

\beq
\left(\begin{array}{c} n_{\nu_e}(t) \\ n_{\nu_\mu}(t) \\
n_{\nu_\tau}(t)\end{array}\right)= \left(\begin{array}{c} a_1 \\
 a_2 \\ a_3 \end{array}\right)
+ \left(\begin{array}{c} b_1 \\ b_2 \\ b_3 \end{array}\right)
\ e^{-t/T_1}
+ \left(\begin{array}{c} c_1 \\  c_2 \\
 c_3 \end{array}\right) \ e^{-t/T_2}\,,\label{3}\eeq

    with
    \beq \sum_{i=1}^3 a_i=1,\quad \sum_{i=1}^3 b_i=0,\quad\sum_{i=1}^3 c_i=0\,. \eeq

    Note, that in relaxation model the neutrino flavour lepton numbers are violated,
while the total sum of the numbers are constant (also as in the
case of oscillation model).

    To illustrate the possibilities of the relaxation model the particular
"simple" solution can be found under the following assumptions:

a) $W_{e \mu}=W_{\mu e},\,  W_{e \tau}=W_{\tau e},\, W_{\mu
\tau}=W_{\tau \mu}$,

b) $W_{\mu \tau} \gg W_{e \mu},\,W_{e \tau}$

    Now equation  (\ref{3}) looks as:

\beq
\left(\begin{array}{c} n_{\nu_e}(t) \\ n_{\nu_\mu}(t) \\
n_{\nu_\tau}(t)\end{array}\right)= \left(\begin{array}{c} 1/3 \\
1/3 \\ 1/3\end{array}\right)
+b\left(\begin{array}{c} 0 \\ 1\\ -1\end{array}\right)
\ e^{-2W_{\mu \tau}t}
+c\left(\begin{array}{c} -1 \\ 1/2 \\
 1/2\end{array}\right) \ e^{-3/2(W_{e \mu}+W_{e
\tau})t}\,.\label{5}\eeq

    For small mass differences between
different neutrino flavours ($m_{\nu_e}\approx m_{\nu_\mu}\approx
m_{\nu_\tau}$ and $\Delta m_\nu\ll m_\nu$) with
$\gamma=E_\nu/m_\nu \gg 1$ the probability $P$ of observing the
neutrino flavour with energy $E_\nu$ at distance $r$ from
the source is given by:

\beq
\left(\begin{array}{c} P_{\nu_e}(r) \\ P_{\nu_\mu}(r) \\
P_{\nu_\tau}(r)\end{array}\right)= \left(\begin{array}{c} 1/3 \\
 1/3 \\ 1/3\end{array}\right)
+b\left(\begin{array}{c} 0 \\ 1\\ -1\end{array}\right)
\ e^{-r/(\Lambda_0 E_\nu)}
+c\left(\begin{array}{c} -1 \\ 1/2\\
 1/2\end{array}\right) \ e^{-r/(\Lambda_1 E_\nu)}\,.\label{6}\eeq

where we use the new notation $\Lambda_0=(2W_{\mu \tau}m_\nu)^{-1}$

and $\Lambda_1=2/3[(W_{e \mu}+W_{e \tau})m_\nu]^{-1}$.

    Equation (\ref{6}) can be solved for different initial experimental
conditions of neutrino flavour fluxes.

    For initial pure electron neutrino or antineutrino flux
(like in the case of solar or reactor experiments)
$P_{\nu_e}(0)=1$, $P_{\nu_\mu}(0)=P_{\nu_\tau}(0)=0$ the solution
is given by:
\begin{eqnarray}
    P_{\nu_e}(r)&= &1/3[1+2 \ e^{-r/(\Lambda_1 E_\nu)}]\,,\nonumber\\
P_{\nu_\mu}(r)&=& 1/3[1-\ e^{-r/(\Lambda_1 E_\nu)}]\,,\nonumber\\
P_{\nu_\tau}(r)&=& 1/3[1-\ e^{-r/(\Lambda_1 E_\nu)}]\,.
\end{eqnarray}

    From Kamland experiment data \cite {2} it is possible to estimate
the value of $\Lambda_1$ parameter. Taking the effective energy of
reactor antineutrino flux to be equal to 4.8  MeV (here we take
into account the threshold of Kamland detector
$E_{\widetilde{\nu_e}} >$ 3.4  MeV), mean reactor distance 180 km
and $R= P_{\widetilde{\nu_e}\, {\rm
(measured)}}/P_{\widetilde{\nu_e}\,{\rm (expected)}}=0.61$ we
obtain
$$
\Lambda_1 = 43 \!\ km/MeV\,.
$$

    Fig.1 illustrates the distance dependence of reactor
antineutrino fluxes ( for $E_{\widetilde{\nu}}=4.8 \,MeV$).

\begin{figure}[h!]
\includegraphics [width=0.8\textwidth]{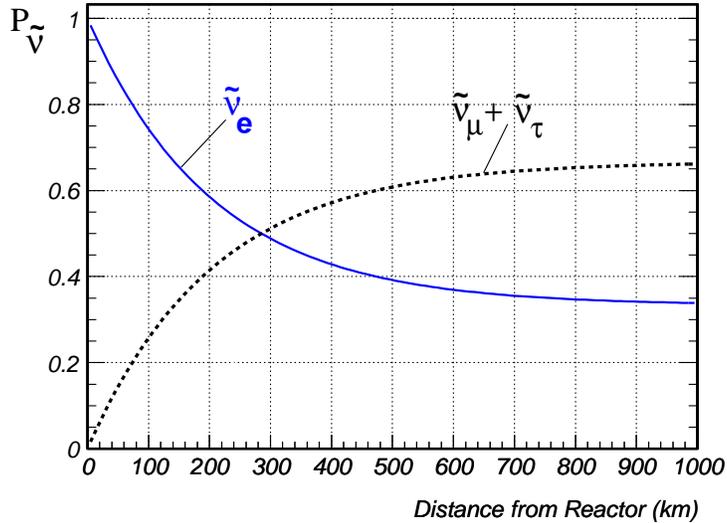}
\vspace*{-5mm} \caption{Probability to observe different flavours
from reactor for $E_{\widetilde{\nu}}=4.8 \,MeV$.}
\end{figure}

    Note, that the final flavour survival probability 1/3 is consistent with the
SNO \cite {1} experimental data with the ratio of measured neutrino flavour fluxes

\beq
 \frac {CC}{NC}
    = 0.306 \pm 0.026({\rm stat}) \pm 0.024({\rm syst})\,,
\eeq

and with the absence of distortion in the measured neutrino spectrum
at low energies, while such distortion is predicted by LMA solution
\cite {11} .

It agrees also with the Homestake results \cite {12} for the ratio $R$
 of observed and predicted by SSM \cite {13}
neutrino rates: $ R=0.34 \pm 0.03 $.

    The deviation from "simple" relaxation model exists for Gallium
experiments \cite {14} \cite {15} with  $ R=0.553 \pm 0.034 $. But it is worth
to mention, that the $R$ value depends on the accuracy of Standard
Solar Model predictions for low energy fluxes.

    For the case of pure initial muon neutrino or antineutrino
flux (like in accelerator experiments) $P_{\nu_\mu}(0)=1$,
$P_{\nu_e}(0)=P_{\nu_\tau}(0)=0$ the solution of Eq. (9) is:

\begin{eqnarray}
P_{\nu_e}(r)&=& 1/3[1-\ e^{-r/(\Lambda_1 E_\nu)}]\,,\nonumber\\
P_{\nu_\mu}(r)&=& 1/3+1/2 \ e^{-r/(\Lambda_0
E_\nu)}+1/6 \ e^{-r/(\Lambda_1 E_\nu)}\,,\nonumber \\
P_{\nu_\tau}(r)&=& 1/3-1/2 \ e^{-r/(\Lambda_0
E_\nu)}+1/6 \ e^{-r/(\Lambda_1 E_\nu)}\,.
\end{eqnarray}

    Preliminary estimation of $\Lambda_ 0$  parameter can be done using K2K
accelerator experiment data \cite {5}. Taking the mean energy of muon neutrino flux
$E_\nu=1.3\,GeV$, distance $r=250\,km$, $R= P_{\nu_\mu\,{\rm
(measured)}}/P_{\nu_\mu\,{\rm (expected)}}=0.70$ we obtain:
$$
\Lambda_0=0.21 \!\ km/MeV\,.
$$

    Fig.2 illustrates the distance dependence of neutrino fluxes
from muon neutrino beam ($E_\nu=1.3\, GeV$).

\begin{figure}[h!]
\includegraphics [width=0.8\textwidth]{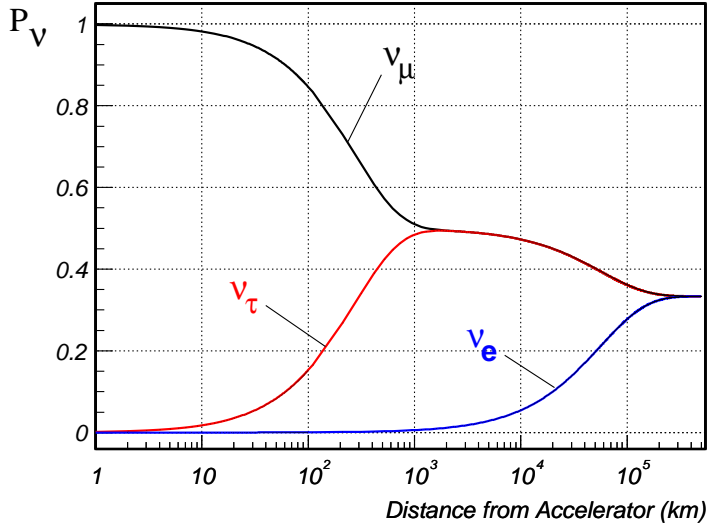}
\vspace*{-5mm} \caption{Probability to observe different flavours
from accelerator muon neutrino beam for $E_{\nu}=1.3\, GeV$.}
\end{figure}

    It is necessary to note that the estimations of neutrino
relaxation parameters $\Lambda_0$ and $\Lambda_1$ are very
preliminary and were made mainly for illustration of
relaxation model. More precise calculations can be done in future,
using exact data of different experiments (taking
also into account atmospheric neutrino data).

    Finally we would like to emphasize the main difference
between the standard oscillation theory and the proposed here
relaxation model: the dependence of neutrino flavour fluxes
from distance is described by the sum of constant and 2 exponents,
instead of oscillation case. Fig.3 illustrates this difference
for reactor antineutrino experiments.

\begin{figure}[h!]
\includegraphics [width=0.8\textwidth]{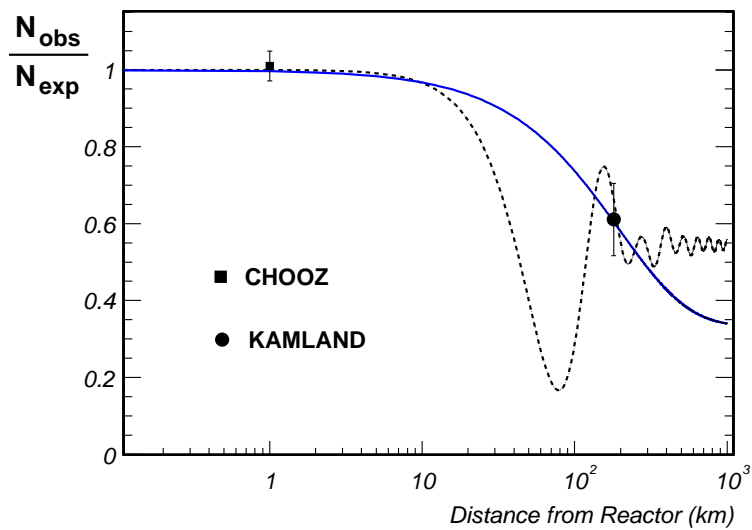}
\vspace*{-5mm} \caption{The ratio of measured to expected
${\widetilde \nu_e}$ flux from  Kamland \cite {2} and Chooz \cite{16}
reactor experiments. Dotted curve --- predictions of oscillation
model with $\sin^2 2\theta =0.91$ and  $\delta m^2=6.9\cdot
10^{-5} (eV)^2$ best fit parameters from \cite {2}. Solid curve ---
predictions of "simple" relaxation  model.}
\end{figure}

    Future reactor and accelerator neutrino experiments
can provide the necessary data to choose between the
neutrino oscillations and the proposed flavour relaxation
model. Of course, the possibility to have a mixture of neutrino
oscillation and relaxation models also exists.

\vspace {0.5cm}

\textsl{Submitted to JETP Letters.}

{}

\end{document}